\documentclass[pra,amsmath,amssymb,twocolumn,showpacs]{revtex4}
\usepackage{mathrsfs}
\usepackage{graphicx}
\usepackage{dcolumn}
\usepackage{epstopdf}
\usepackage{soul}
\usepackage{bm}

\begin{document}


\title{Generation and efficient measurement of single photons from fixed frequency superconducting qubits}


\author{William F. Kindel}
\email[]{william.kindel@colorado.edu}
\affiliation{JILA, University of Colorado and NIST, Boulder, Colorado 80309, USA}
\affiliation{Department of Physics, University of Colorado, Boulder, Colorado 80309, USA}

\author{M. D. Schroer}
\affiliation{GE Aviation, Cincinnati, Ohio 45215, USA}

\author{K. W. Lehnert}
\affiliation{JILA, University of Colorado and NIST, Boulder, Colorado 80309, USA}
\affiliation{Department of Physics, University of Colorado, Boulder, Colorado 80309, USA}
\affiliation{National Institute of Standards and Technology (NIST), Boulder, Colorado 80305, USA}

\date{\today}

\begin{abstract}
We demonstrate and evaluate an on-demand source of single itinerant microwave photons. Photons are generated using a highly coherent, fixed-frequency qubit-cavity system, and a protocol where the microwave control field is far detuned from the photon emission frequency. By using a Josephson parametric amplifier (JPA), we perform efficient single-quadrature detection of the state emerging from the cavity. We characterize the imperfections of the photon generation and detection, including detection inefficiency and state infidelity caused by measurement backaction over a range of JPA gains from 17 to 33 dB. We observe that both detection efficiency and undesirable backaction increase with JPA gain. We find that the density matrix has its maximum single photon component $\rho_{11} = 0.36 \pm 0.01$ at 29 dB JPA gain. At this gain, backaction of the JPA creates cavity photon number fluctuations that we model as a thermal distribution with an average photon number $\bar{n} = 0.041 \pm 0.003$.
\end{abstract}




\maketitle

\section{Introduction}

With the advent of superconducting circuits used as quantum technologies, the prospect of processing quantum information has become less remote. In particular, the circuit quantum electrodynamics (CQED) concept has seen many recent successes. This architecture has been used to prepare single mode states in arbitrary Fock~\cite{Hofheinz:2009} and Schrodinger cat states~\cite{Kirchmair:2013}, entangle and teleport states~\cite{Steffen:2013}, and perform multiqubit gates with high fidelity~\cite{Barends:2014}. An important challenge is creating networks of CQED systems either to create a quantum communication network or as an architecture for building a scalable quantum information processor~\cite{Kimble:2008}. As a CQED system comprises qubits in a microwave resonant circuit (or cavity)~\cite{Wallraff:2004}, communication among CQED modules is naturally accomplished by traveling or itinerant microwave fields coupled to those resonant modes~\cite{Wenner:2014,Flurin:2015}.

Exploiting itinerant modes for communication among various CQED modules requires transferring quantum information between stationary and propagating modes. To that end, there have been a number of efforts to generate and analyze propagating non-classical states using qubit based CQED systems. Initial work generated states using control pulses at the same frequency as the emitted photons~\cite{Houck:2007}. Subsequent work overcame this undesirable aspect by using superconducting qubits with rapidly tunable resonance frequencies~\cite{Eichler:2011,Yin:2013}. More recently, higher level qubit transitions were used to create itinerant states where the multiple control fields were detuned from the emitted state~\cite{Pechal:2014}.

Integrating propagating quantum states with low bandwidth modules places further demand on the systems generating the state. Low bandwidth modules, such as electro-mechanical devices, have been used to capture, store, and release microwave signals~\cite{Palomaki:2013n,Palomaki:2013s}, and they may provide a quantum interface between microwave and optical signals~\cite{Andrews:2014}. However, to create propagating modes with a spectra narrow enough to be compatible with these signal processing elements requires highly coherent CQED devices. The most coherent systems are currently transmon qubits coupled to a 3D cavity where neither the qubit nor the cavity is tunable~\cite{Paik:2011,Schoelkopf:2013}. Consequently, the only controls available for creating itinerant non-classical states are microwave fields.  Furthermore, these control fields should be far detuned from the emitted quantum state to avoid interfering with that state.

In this letter, we show the creation of single itinerant photons. These photons are generated from a fixed frequency transmon qubit coupled to a 3D cavity by an off resonant control field. We efficiently measure and perform tomography on single photons using a Josephson parametric amplifier (JPA). We maximize the single photon component over a range of JPA gains from 17 to 33~dB finding an optimum JPA gain of 29~dB. Here, we measure a state whose single photon component is $\rho_{11} =$ 0.361 with $\pm$~0.005 statistical and $\pm$~0.005 systematic uncertainties. At this gain, the JPA backaction creates photon number fluctuations that we model as a thermal distribution with an average photon number $\bar{n} = 0.041 \pm 0.003$. We characterize the limitations of the generation and detection including amplifier noise and backaction on the qubit-cavity system. Accounting for both amplifier backaction and for cavity decay to unmeasured ports we calculate the expected mixed state exiting the cavity. Within uncertainties, our expected state is consistent with our measurements.

\section{Photon Generation}\label{sec:PhotGen}

\begin{figure} 
\includegraphics[width=3.375in]{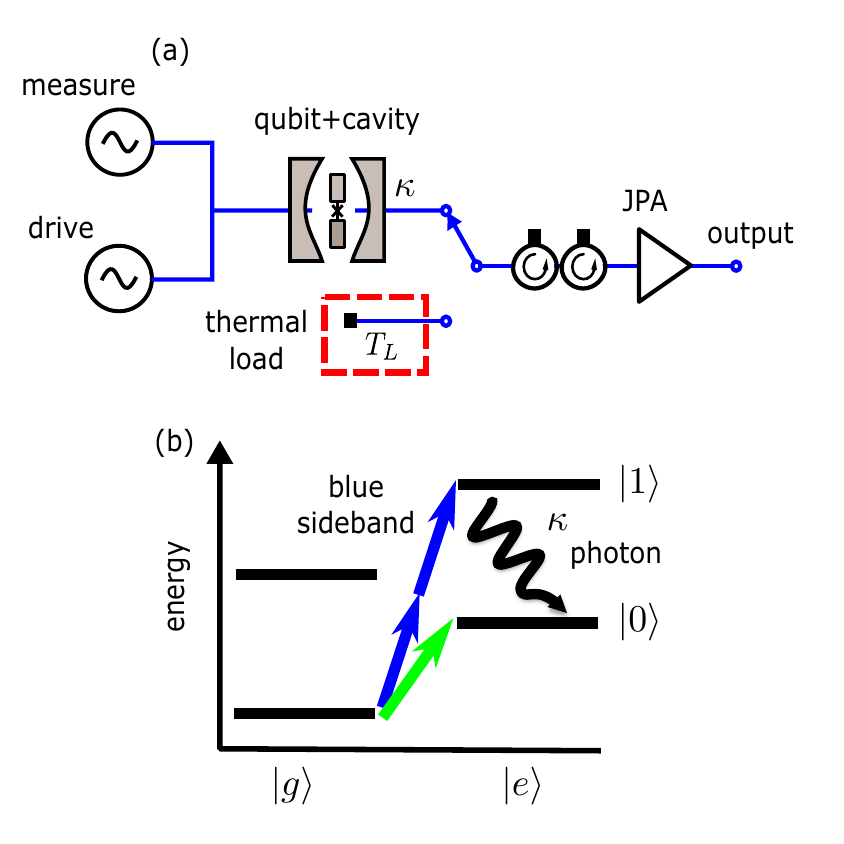}%
\caption{Simplified schematic and energy level diagram for photon generation. (a) The drive and measurement microwave generators couple to the input of the qubit cavity system where the measurement tone excites the cavity in order to infer the qubit state, and the drive tone manipulates the state of the qubit-cavity system. A switch can connect the amplifier chain to a thermal noise source allowing for an independent characterization of the measurement efficiency.  (b) Starting with the qubit in the ground state and zero photons in the cavity, a two photon blue side-band pulse (blue double arrow) excites the qubit and creates one photon. The photon then decays out of the system, creating a propagating photon in the microwave lines. We compare this process with a control, where a qubit pulse (green arrow) directly excites the qubit and no photon is created.}\label{fig:cartoon}
\end{figure}

We generate single itinerant photons by first creating stationary single photons in a qubit-cavity system and then letting them decay through a strongly coupled output mode. In Fig. \ref{fig:cartoon} we depict a simplified schematic and the protocol for generating propagating photons using only cavity control fields. We work with a CQED system in the strong dispersive regime with the qubit frequency $\omega_q$/2$\pi =3.495$~GHz and the cavity frequency $\omega_c$/2$\pi = 5.804$~GHz. In this regime, the cavity shifts by $2\chi$/2$\pi = -2.0$~MHz depending on the state of the qubit.  A microwave tone at frequency $\omega_b \approx (\omega_c + \omega_q)/2$  drives transitions between the state $|g,0\rangle$ and $|e,1\rangle$, where we label the states of the system as $|$qubit state, cavity photon number$\rangle$. Starting in $|g,0\rangle$, a $\pi$-pulse on this two-photon blue-sideband transition prepares the system in the $|e,1\rangle$ state~\cite{Blais:2007,Leek:2009}. Making use of the long $T_1 =10$~$\mu$s time of transmon qubits, we arrange for the dominant relaxation mechanism from this cavity state to be emission of a microwave photon (at frequency $\omega_c$) by coupling the cavity to a transmission line with energy decay rate $\kappa_{out}/2\pi = 300$~kHz.

\begin{figure} 
\includegraphics[width=3.375in]{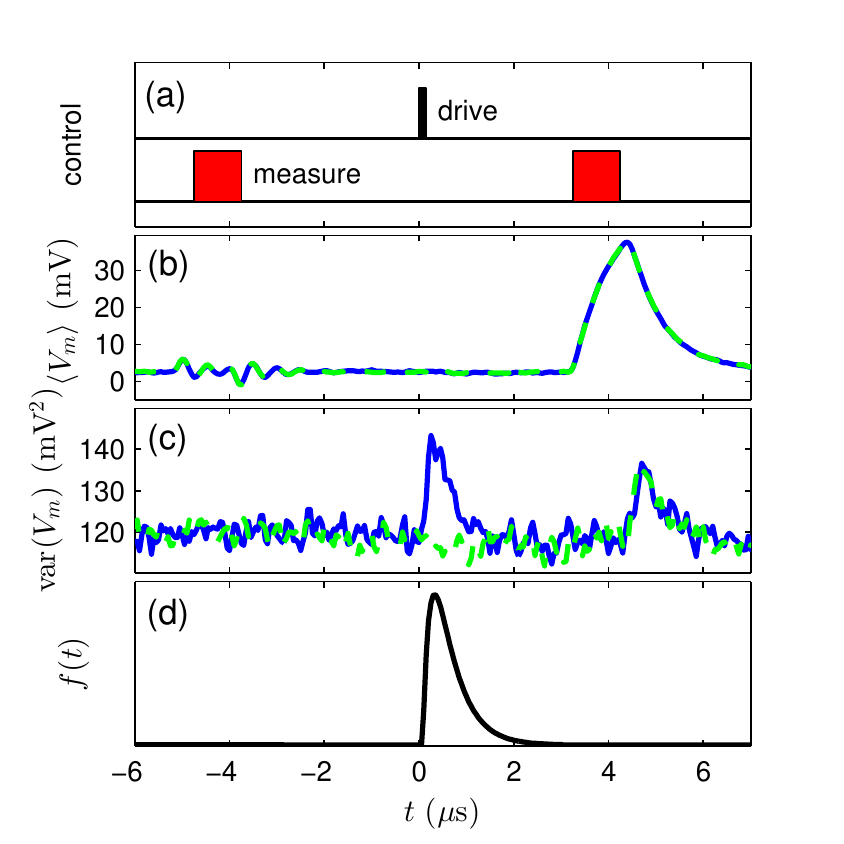}
\caption{Time domain depiction of photon generation and detection. (a) The figures shows a timing diagram for the photon creation and measurement sequence.  The drive tone (black) either creates a photon or excites the qubit for the  vacuum calibration.  Qubit readout tones (red) are used before and after the drive. (b) The plot shows the mean voltage $\langle V_{m} \rangle (t)$ measured  from 7000 individual time traces with the drive at the blue sideband frequency (solid blue) or at the qubit's frequency (dashed green). (c) Plotted is the variance of the individual measurements $\mathrm{var}( V_{m} ) (t)$ with the drive at the blue sideband frequency (solid blue) and at the qubit frequency (dashed green).  In blue, the photon power can be seen following the blue sideband pulse at $t = 0$~$\mu$s. For the control in green, no such signal is seen.  (d) The mode matching function $f(t)$  is applied to each individual voltage trace, $V_{m}(t)$, to extract the quadrature measurement for the propagation state exciting the cavity.}\label{fig:pulse}
\end{figure}

To increase the fidelity of the photon generation, we implement a pulse sequence that measures and conditions on the state of the qubit. The measurement of the qubit state occurs both before and after the blue sideband drive pulse, as shown in the timing diagram in Fig. 2(a). By selecting the trials in which pre-pulse measurement indicates that the qubit is in its ground state, we eliminate most of the approximately 6\% of trials in which the qubit begins in the excited state~\cite{Vijay:2011,Riste:2012}.  By measuring the qubit state after the pulse and cavity decay, we select those trials in which we find the qubit in the excited state. The second selection eliminates those cases ($\approx$ 26\%) in which the qubit decays during the photon emission process or in which the protocol failed due to pulse infidelity. (If the qubit decays during the photon emission process, the photon is emitted at a different frequency.)

To provide a calibration for our photon measurements, we modify the photon generation protocol by the minimal amount that ensures no photons are generated. We replace the $\pi$-pulse on the blue sideband transition with a $\pi$-pulse on the qubit transition. This control sequence prepares the qubit cavity system in the $|e,0\rangle$ state rather than the $|e,1\rangle$ state; thus, the cavity cannot emit a photon. All other aspects, including the qubit state selection and the data processing procedure are common to both protocols.

Because the qubit measurement must preserve the qubit in its ground state for photon creation, the measurement must have a quantum nondemolition (QND) character. Furthermore, because qubit readout occurs both before and after photon detection takes place, the qubit readout must be compatible with photon measurement. To satisfy these two requirements, we use a slightly modified version of the dispersive,  JPA-backed qubit readout, which detects the qubit state dependent shift of the microwave cavity's resonance frequency~\cite{Vijay:2011,Riste:2012,Vijay:2012}. In contrast to Ref.~\cite{Riste:2012}, we operate the JPA with its narrow-band gain centered on the qubit-excited cavity resonance frequency ($\omega_c + \chi$). In our configuration, a single pump field, from which the JPA derives its phase dependent gain, would be resonant with the cavity. We circumvent this problem by using the so-called double pump scheme, where two pumps tuned symmetrically above and below the JPA center frequency provide power to the JPA~\cite{Kamal:2009}. We generate the two pumps by amplitude modulating a carrier tone, where the phase of the carrier tone determines which quadrature the JPA amplifies. To detect the state of the qubit, we excite the cavity with a measurement tone at frequency ($\omega_c + \chi$), then one quadrature of the transmitted field is amplified by the JPA and other amplifiers before being mixed down using the carrier tone as the mixer's local oscillator. The output of this measurement chain is one voltage trace $V_m(t)$ for each iteration of the protocol. In Fig.~\ref{fig:pulse}(b) we show the average of approximately 7000 such iterations $\langle V_m \rangle(t)$, illustrating that the transmitted \textit{amplitude} of the cavity's measurement tone can discriminate between the two qubit states.

\section{Photon Detection}

To characterize the propagating state, we perform tomography on the output mode of the cavity to determine its density matrix $\bm{\rho}_m$. Our tomography procedure measures a single quadrature of the cavity output field during the photon generation protocol. In general, the density matrix can be reconstructed by repeating this generation and measurement protocol many times and at several different values of JPA carrier phase. Because we choose to create a single photon, which has a phase independent density matrix, we unlock the phase between the generation pulse and the JPA carrier thus sampling all phases uniformly. From a histogram made from many quadrature measurements we extract the diagonal elements of the density matrix written in the photon number basis. If there were any off-diagonal elements, these would vanish due to the phase averaging.

Because the JPA continuously amplifies the cavity output field, information about the emitted photon's density matrix is present in the measurements of $V_m(t)$.  We form one measurement set by repeating the pulse sequence shown in Fig. \ref{fig:pulse}(a) 7000 times (with the $\pi$-pulse applied to the blue sideband). The temporal envelope of the photon can be seen in the variance of the set $ \mathrm{var}(V_{m})(t)$ at time $t=0\ \mu$s [Fig.~\ref{fig:pulse}(c)]. As expected for a diagonal density matrix, no feature is present in $\langle V_m \rangle(t)$ at $t=0$. 

\begin{figure} 
\includegraphics[width=3.375in]{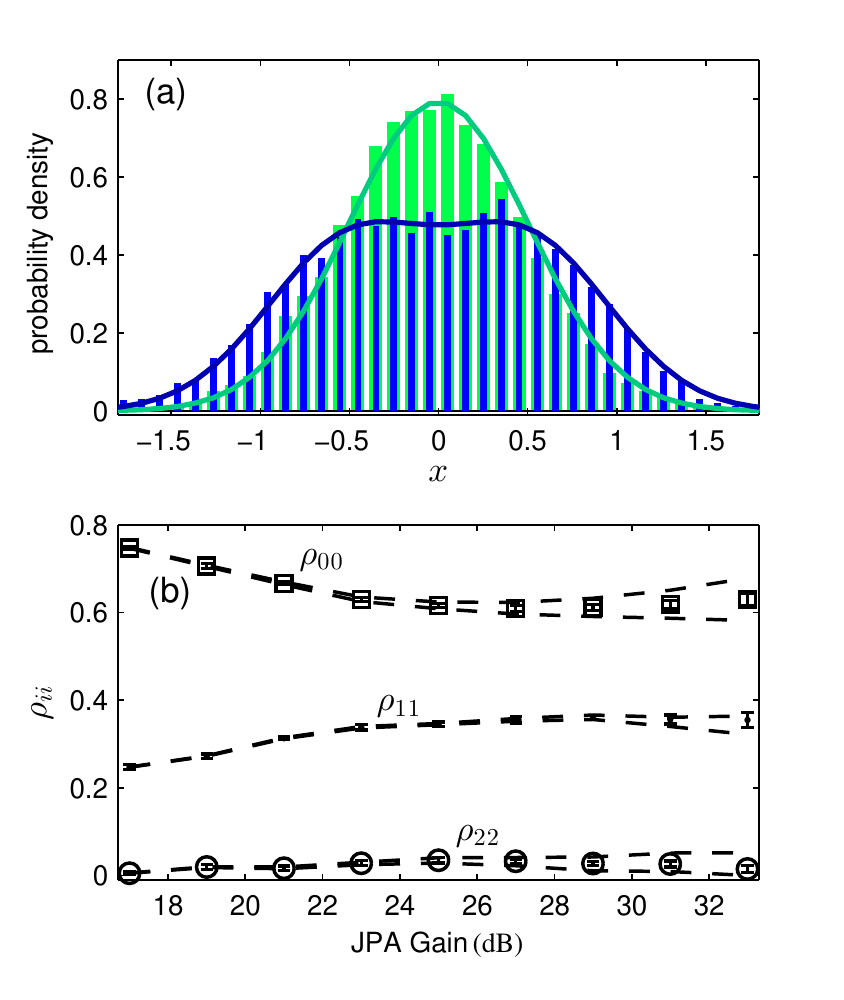}
\caption{Photon state tomography. (a) Histogram of a quadrature measurement set of single photons (narrow blue bars) and the no-photon control (wide green bars) with the JPA gain at 27 dB. The histograms are plotted as a probability density (bars) and are fit by the expected distribution for a diagonal density matrix with a 3 photon Fock basis (solid lines). (b) Diagonal density matrix elements determined from fitting the quadrature histogram over a range of JPA gains from 17 dB to 33 dB. The error bars are determined from the standard deviation of the mean using 8 sets of the quadrature measurements at each gain; (only 4 sets are used at 17 and 33 dB). The systemic uncertainties are indicated with dashed lines. }\label{fig:rho}
\end{figure}

To complete the reconstruction, each $V_{m}(t)$ must be mapped to a single quadrature value of the propagating mode (Appendix~\ref{AppenCal}). For each trace we assign an uncalibrated quadrature value $V_q$ by minimizing 
\begin{equation}
\mathcal{J} = \int \left[V_m(t)-(V_q f(t) + \bar{b})\right]^2 dt \label{eq:error}
\end{equation}
over $V_q$ where $\bar{b}$ is a measurement of the background voltage. The mode matching function $f(t)$ [Fig.~\ref{fig:pulse}(d)] is optimized experimentally by matching $f(t)^2$ to data similar to what is shown in Fig.~\ref{fig:pulse}(c). A histogram of this set of measurements is shown as the narrow blue bars in Fig.~\ref{fig:rho}(a). We generate a calibration data set following the same procedure as was used to generate the measurement set, but using the control protocol ($\pi$-pulse on the qubit transition). Indeed, when no photon is created, no extra variance is visible in $\mathrm{var}(V_m)(t)$ at $t=0$ [Fig.~\ref{fig:pulse}(c)]. Reducing each trace to quadrature values, we find the histogram shown in Fig.~\ref{fig:rho}(a).

To calibrate the quadrature values, we fit the histograms of the calibration data sets to Gaussian distributions using the gain of the full measurement apparatus as the only fit parameter. But due to the fact that a small fraction of the JPA output is injected back into the cavity through the finite isolation of the circulators (determined in Sec.~\ref{sec:char}), we do not assume that the cavity is in its vacuum state. Rather, we expect to prepare the cavity in a mixed, squeezed state with $\bar{n} \ll 1$. Although the large isolation of the circulators ensures that the cavity's squeezed quadrature has almost exactly vacuum variance, the amplified quadrature can have variance measurably larger than vacuum, particularly at larger values of JPA gain. Because there is a fixed, but unknown phase relationship between the quadrature we measure and the squeezed quadrature of the cavity, we calibrate assuming we measure the squeezed quadrature and assuming we measure the amplified quadrature. We use these two cases to bound the systematic uncertainty in our calibration of the vacuum variance, where we use the convention that one quadrature of a vacuum state has variance ${\rm var}(x) = 1/4$.   

Finally, we complete the tomography by fitting the histograms to probability distributions for diagonal elements of the density matrix in a 3 photon basis yielding one measurement of the density matrix. (A 3 photon basis is sufficient because the three photon component $ \rho_{33}$ is indistinguishable from 0.) Each density matrix element is determined as the average value of multiple realizations of each measurement set. To optimize the photon measurement we find the density matrix elements over a range of JPA gains from 17 to 33~dB [Fig.~\ref{fig:rho}(b)]. These density matrix elements are sensitive to all imperfection in the photon generation and detection. For ideal generation and detection, the $\rho_{11}$ component would be 1 and all others 0. Instead, the single photon component has a peak value of $\rho_{11} =$ 0.361 with $\pm$~0.005 statistical and $\pm$~0.005 systematic uncertainties at a JPA gain of 29~dB. However, at this gain there is a two photon component of $\rho_{22} =$ 0.027 with $\pm$ 0.005 statistical and $\pm$~0.015 systematic uncertainties. Considering the two photon generation relative to single photon generation, we have $2\rho_{22}/\rho_{11} \approx g_2(0) = 0.32 \pm 0.07 $ with a 0.15 to 0.41 systematic uncertainty bound. [For comparison, $g_2(0) = 1$ for any coherent state.] By decreasing the JPA gain to 17~dB, the two photon component becomes $\rho_{22} =$  0.005 $\pm$ 0.003  with a single photon component dropping to $\rho_{11} =$0.247 $\pm$ 0.004, giving $g_2(0) =  0.15$~$\pm$ 0.08. These results show that an increase in the JPA gain improves the measurement efficiency, but also increases the measurement backaction. The deleterious effects of this backaction can be seen in both the larger systematic uncertainty at larger JPA gain and the increased probability of creating two photons instead of just one.

\section{Characterizing generation and measurement}\label{sec:char}

To understand the limitations of single photon generation and detection, we characterize the experimental imperfections over a range of JPA gains.  In particular, we independently quantify the measurement backaction and characterize the internal loss of the cavity and measurement efficiency. This characterization gives us a prediction for the state that we create and an expectation for how efficiently we can measure it. We compare our expectation with our measurements to validate our understanding of the photon generation process. 

\begin{figure} 
\includegraphics[width=3.375in]{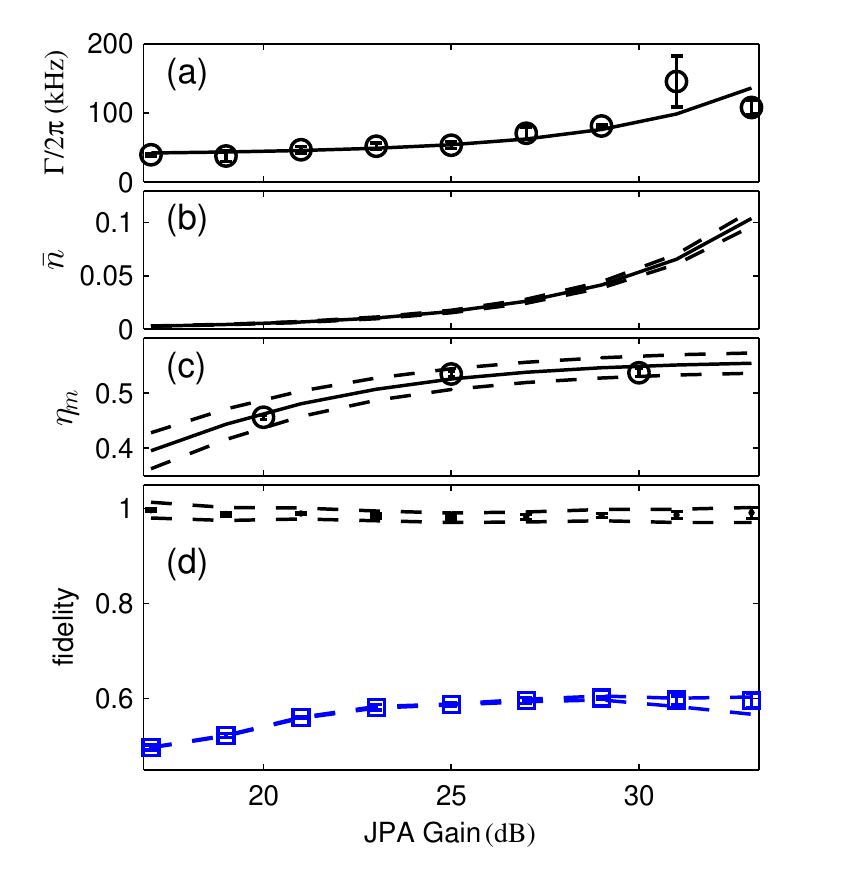} 
\caption{Characterizing and correcting for imperfections in generation and detection. (a) The qubit dephasing rate ($\Gamma = 1/T_2^*$) is shown (points) over a range of JPA gains along with a fit (line) to Eq.~\ref{eq:dephase} over the range of JPA gains (main text). (b) The solid line shows the best estimate of the average photon number in the cavity $\bar{n}$ (uncertainty in dashed lines) due to the JPA at each gain used in the experiment. (c) The measurement efficiency $\eta_{m}$ is determined from a thermal sweep at three JPA gains (circles). This quantity is interpolated over the range of JPA gains by fitting the data points to Eq.~\ref{eq:Nadd} (Appendix~\ref{AppenThm}) with the uncertainty in dashed lines. (d) The plots show the fidelity of the measured density matrix with respect to: an ideal single photon (blue squares), and the density matrix we expect to measure given pure photon generation in the cavity (black points). Here, the statistical one-standard-deviation uncertainties are plotted as errors bars. The systematic error plotted in dashed lines is calculated from both the systematic uncertainty in $\eta_{m}$ and in the density matrix elements.}\label{fig:results}
\end{figure}

To quantify the cavity photon variance due to JPA backaction, we study the qubit dephasing~\cite{Murch:2013}. As seen in the dispersive Hamiltonian for the qubit cavity system in Eq.~\ref{eq:dispersive}, the frequency of the qubit depends on the number of photons in the cavity. Therefore, a varying number of cavity photons will dephase the qubit. Because the qubit resonance frequency is far-detuned from the cavity, we treat the squeezed cavity field as dephasing the qubit with a thermal distribution. With this assumption, we calculate a photon dephasing rate~\cite{Sears:2012}    
\begin{equation}
\Gamma = \Gamma_{0}  +\kappa [2 \bar{n} + 2 \bar{n}^2 + O(\bar{n}^4) ]   \label{eq:dephase}
\end{equation} 
in terms of the average number of cavity photons $\bar{n}$, cavity decay rate $\kappa$, and the intrinsic dephasing rate $\Gamma_{0}$. Measurements of the qubit dephasing rate ($\Gamma = 1/T_2^*$) over the range of JPA gains are shown in Fig.~\ref{fig:results}(a).

We assume that $\bar{n}$ follows a model characterized by a single isolation parameter $L$, which characterizes the fraction of JPA output misdirected into the cavity as
\begin{equation}
\bar{n} = (1/4)L(G_{\mathrm{JPA}} -1), \label{eq:nbar}
\end{equation}
where $G_{\mathrm{JPA}}$ is the JPA gain. We fit this model substituted into Eq.~\ref{eq:dephase} to measurements of $\Gamma$, extracting $L = (2.1 \pm 0.1) \times 10^{-4}$ (-37 dB) and $\Gamma_0/2\pi = 40 \pm 2 $~kHz [Fig.~\ref{fig:results}(a)]. The isolation is consistent with the specifications for the two commercial circulators between the cavity and the JPA [Fig. 1(a)]. From this model, we find the average intercavity photon number due to JPA gain [Fig.~\ref{fig:results}(b)]. At our peak $\rho_{11}$ (29~dB of JPA gain), there is an average photon $\bar{n} = 0.041 \pm 0.003$ in the cavity due to JPA backaction, likely accounting for most of the two photon component in the measured density matrix.

We make a prediction of the state we expect to measure if the photon generation protocol created a pure single photon state in the cavity. First, we form an expression for the density matrix of the output mode $\bm{\rho}_{\mathrm{out}}$ by accounting for the coupling of the cavity field to unmeasured ports. Due to the relative coupling rates, a cavity photon has a $\kappa_{out}/\kappa$ probability of decaying to the output port, where the coupling rate to the output port is $\kappa_{out}/2\pi = $~300~kHz and the total decay rate is $\kappa/2\pi =$~410~kHz. We would therefore expect to generate a propagating state $\bm{\rho}_{\mathrm{out}} $ characterized by
\begin{equation}
\bm{\rho}_{\mathrm{out}} = \frac{\kappa-\kappa_{out}}{\kappa}|0 \rangle \langle  0| + \frac{\kappa_{out}}{\kappa} |1 \rangle \langle  1|. \label{eq:rhoout}
\end{equation}

Next, we form an expectation for how well we can measure $\bm{\rho}_{\mathrm{out}}$ by independently characterizing the measurement inefficiency. We cast this expectation as the density matrix $\bm{\rho}_{\mathrm{exp}}$. We determine the efficiency $\eta_m$ by injecting states of known variance [generated by the thermal load in Fig.~\ref{fig:cartoon}(a)] into the measurement chain. By adjusting the input variance (thermal load temperature), we can determine the additional variance introduced by the measurement, and therefore the efficiency (Appendix~\ref{AppenThm}). This procedure is performed at three JPA gains shown in circles in Fig.~\ref{fig:results}(c). The efficiencies are interpolated over the range of JPA gains (solid line). In terms of $\eta_m$, our expectation for the density matrix we should measure for pure cavity photon generation is
\begin{equation}
\bm{\rho}_{\mathrm{exp}} = (1 - \frac{\kappa_{out}}{\kappa}\eta_m)|0 \rangle \langle  0| + \frac{\kappa_{out}}{\kappa} \eta_m |1 \rangle \langle  1|. \label{eq:rhoexp}
\end{equation}    

Finally, we compare our measurement to expectation by computing the fidelity~\footnote{The fidelity of a state A ($\bm{\rho}_A$) with a state B ($\bm{\rho}_A$) is $\mathcal{F} =\mathrm{tr}\left( \sqrt{\sqrt{\bm{\rho}_B}\bm{\rho}_A \sqrt{\bm{\rho}_B}} \right)$.} of $\bm{\rho}_m$ with respect to $\bm{\rho}_{\mathrm{exp}}$. We find that they are identical (unit fidelity) within uncertainty [Fig.~4(d)]. This agreement shows that we are able to accurately and independently characterize the measurement inefficiency and undesired cavity loss. (The measured two-photon component contributes negligibly to the infidelity.) For comparison, we compute the fidelity of the measured density matrix $\bm{\rho}_m$ with respect to the density matrix of a single photon [Fig.~4(d)], which quantifies our combined ability to generate and detect single photons. This fidelity has a peak value of $\mathcal{F} = 0.600\pm  0.008$  also at 29~dB JPA gain.

\section{Conclusions}

The protocol we have demonstrated for generating single microwave photons on demand is well suited for transferring quantum states to narrow bandwidth signal processing modules, such as certain types of electrooptic convertors that are under development~\cite{Andrews:2014}. In particular, the compatibility of the protocol with fixed-frequency, highly coherent qubit-cavity systems ensures that the photons can be emitted into narrow frequency windows. However, our detailed characterization of the protocol reveals an undesirable backaction of the measurement apparatus onto both the qubit and the cavity. Contingent on advances in low loss isolator elements~\cite{Kerckhoff:2015,Sliwa:2015}, it may be possible to mitigate this backaction through the use of more isolation while at the same time reducing losses in detection. Otherwise, a more complicated protocol that pulses on the JPA pumps only when the measurement is desired may minimize the backaction. Thus, the measurements presented here highlight the importance of isolator elements in quantum information processing, and provide a method to transfer information between 3D qubits and itinerant microwave fields.

\appendix
\section{Detailed description of photon generation and detection apparatus}\label{AppenDetails}

\begin{figure*} 
\includegraphics[width=6.75in]{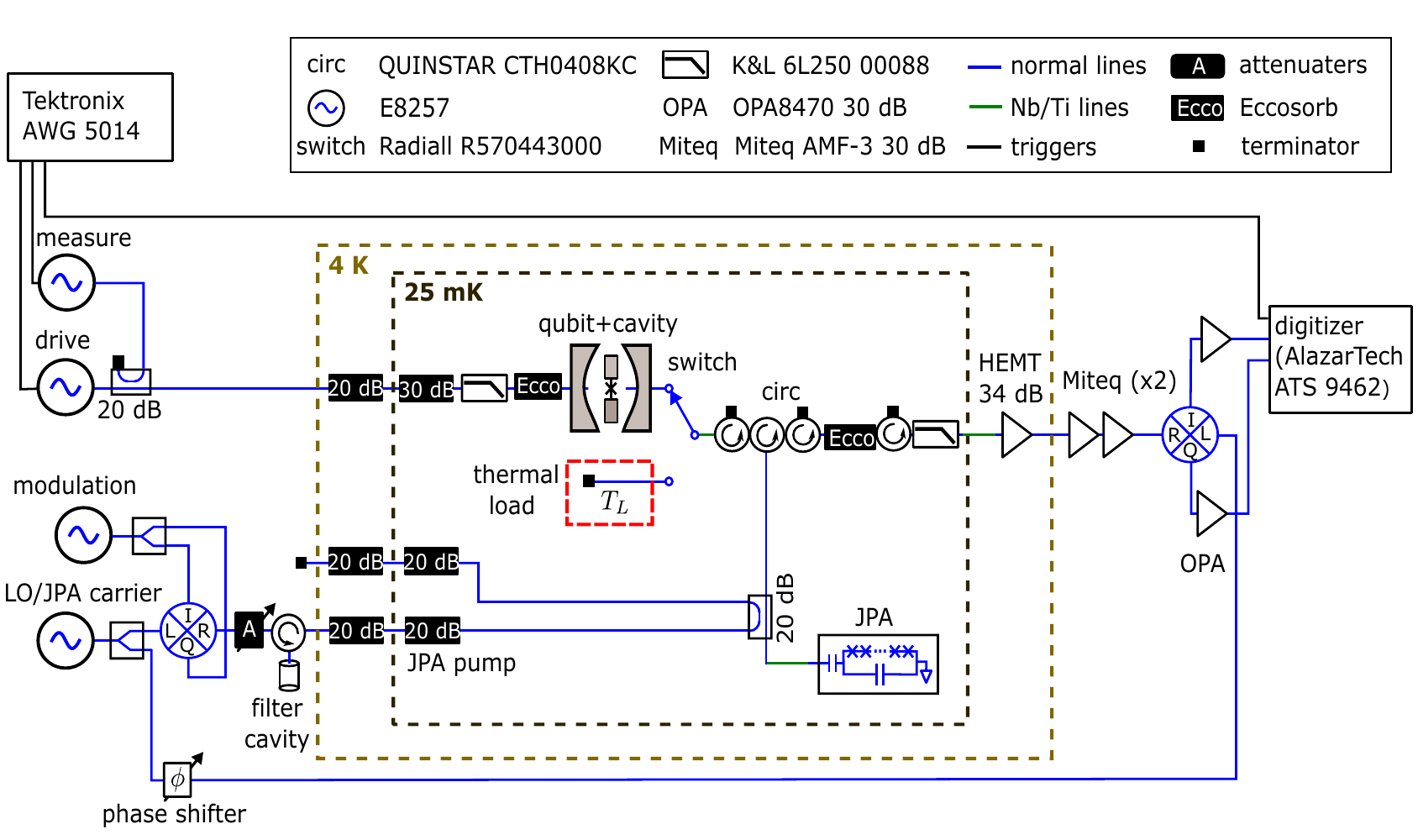}
\caption{The microwave schematic for the experiment.}\label{fig:microwave}
\end{figure*}

The experimental details are summarized by the microwave schematic (Fig.~\ref{fig:microwave}). The experiment is conducted in an Oxford Triton 200 dilution refrigerator with the qubit-cavity system anchored to the (T $<$ 25 mK) base temperature region. The superconducting qubit is a hybrid TiN/Al design fabricated on high resistivity, intrinsic silicon similar to  Ref.~\cite{Chang2013}. The qubit is coupled to a cavity milled from a block of extruded T6061 Aluminum with all surfaces mechanically polished. The measured qubit frequency, cavity frequency, and dispersive shift are $\omega_q$/2$\pi$~=~4.385 GHz, $\omega_c$/2$\pi$~=~5.805 GHz, $\chi/2\pi~=~-1.0$~MHz respectively. These parameters are defined for the dispersive limit
\begin{equation}
H_{dis} = \omega_c a^{\dagger}a +  \omega_q\frac{\sigma_z}{2} + 2 \chi a^{\dagger}a \frac{\sigma_z}{2}    \label{eq:dispersive}
\end{equation}
of the Rabi Hamiltonian. The 410~kHz linewidth of the cavity is dominated by the output coupling ports, $\kappa_{out}/2\pi = 310$~kHz, and the qubit has $T_1$~=~10.2$~\mu$s , and $T_2^*$~=~4.0$~\mu$s. These parameters place the system in the strong dispersive regime. Except for some variation in the qubit and cavity coherence parameters, the qubit-cavity system used in this work and in Ref.~\cite{Schroer:2014} is the same. This system is controlled and measured via injecting tones shown on the left. The strongly coupled output on the right leads to a switch, which either connects the qubit-cavity system or the thermal load to the measurement chain.

After the switch, circulators route signals into the JPA. The JPA is a nonlinear lumped element LC resonator pumped by rf power injected using a 20~dB directional coupler. It is operated with a signal gain from 17 to 33~dB and with an approximate gain bandwidth product of 43~MHz. The JPA is pumped with two tones in the so called double pump method by modulating a 5.806~GHz carrier by 240~MHz using an IQ mixer. The carrier is suppressed by the modulation and is further reduced by a notch cavity filter. The output of the JPA is routed into a HEMT amplifier and room temperature amplifiers before being mixed down by a second IQ mixer.  A copy of the JPA carrier is used as this mixer's local oscillator and phase shifted so that the JPA's amplified quadrature exits the I-port of the mixer. This output is then further amplified and then digitized. An arbitrary waveform generator determines the protocol timing by triggering the drive tone, the measurement's tone, and the digitizer.   

\section{ Individual quadrature measurement}\label{AppenCal}

Each instance of the protocol produces a discretely sampled V$_m(t)$. From this raw data, we make an estimate of one quadrature of the mode emitted by the cavity when the drive pulse is applied. We desire a mode matching function $f(t)$, that weights the time average of $V_m(t)$ to produce an optimum estimate of the quadrature value. For infinite measurement bandwidth, we expect the optimum $f(t)$ to be a decaying exponential pulse with decay constant $\kappa$~\cite{Eichler:2013} and a rise time equal to the duration of the drive pulse (150 ns). Due to the finite JPA bandwidth, we anticipate that $f(t)$ is found by convolving the infinite bandwidth optimum by the JPA impulse response~\cite{Gibbs:2011}. In practice, we write $f(t)$ as a function of three parameters (rise time, decay constant, and JPA bandwidth) and adjust these to minimize the zero-photon contribution of the density matrix extracted from the data set. We perform this determination of $f(t)$ once, using photon creation and calibration data sets that are not used in subsequent analysis. As seen in Fig.~\ref{fig:pulse}(d), $f^2(t)$ looks like the photon creation variance [Fig.~\ref{fig:pulse}(c)], indicating that our optimization of $f(t)$ has produced a sensible result.  
 
Extracting the quadrature value is complicated by the presence of a dc offset in the $V_m(t)$ which drifts during the acquisition of a full data set. To remove drifts in the dc offset from the quadrature measurement, we perform a linear least-squares fit using a measurement of the background voltage in addition to the mode matching function. To do so, we minimize the cost function (Eq.~\ref{eq:error}) for background $\bar{b} = V_{dc} \tilde{b}(t)$, where
 $V_{dc}$ is the dc voltage and $\tilde{b}(t)$ is the windowing function that defines when the dc offset is measured. The windowing function is a piecewise constant function that is nonzero during most of a 56 microsecond interval that includes the photon generation protocol. But during the qubit measurements, $\tilde{b}(t) = 0$. The result of the cost function minimization is an analytic expression for an individual quadrature measurement
\begin{equation}
V_q = \int V_{m}(t) f(t) dt - \int V_{m}(t) \tilde{b}(t) dt \int \tilde{b}(t) f(t) dt. \label{eq:LSF}
\end{equation}
This expression is applied to each $V_{m}(t)$ resulting in a set of uncalibrated quadrature measurements of the propagating mode.

\section{Independent measurement characterization}\label{AppenThm}

\begin{figure}
\includegraphics[width=3.375in]{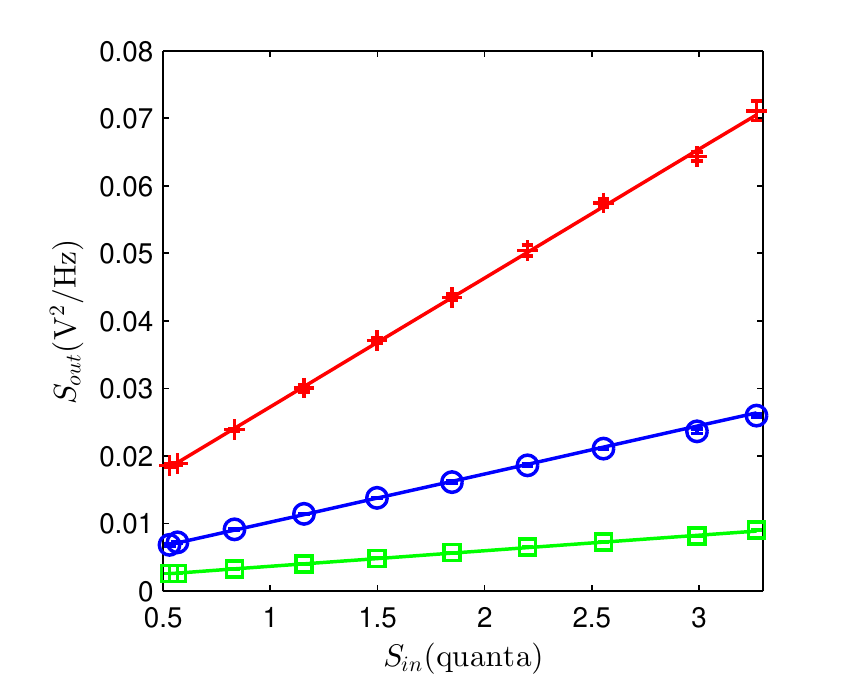}
\caption{Thermal sweep data for the JPA operated at 20 (green squares), 25 (blue circles) and 30~dB (red crosses) gains. The data are fit to Eq. \ref{eq:Sout} (solid lines). The input noise source used in the thermal sweep is a 50~$\Omega$ resister whose temperature is adjusted from 79~mK to 900~mK. This thermal noise power is expressed in units of quanta at 5.8~GHz on the x-axis.}\label{fig:therm}
\end{figure}

In order to independently determine the measurement efficiency, we inject states of known variance into the measurement chain. As we adjust the input variance into the measurement chain, we determine the additional variance $N_{add}$ introduced by the measurement. The source of known variance is a 50~$\Omega$ resistor on a variable temperature stage connected by a switch to the JPA input (Fig.~\ref{fig:microwave}). We perform the determination of $N_{add}$ at three different JPA gains-- 20, 25, and 30~dB (Fig.~\ref{fig:therm}). The data are plotted as output-noise power-spectral-density $S_{out}$ against input power spectral density $S_{in}$. We extract $N_{add}$ by fitting data in Fig.~\ref{fig:therm} to
\begin{equation}
S_{out} = G (S_{in} +  N_{add}),  \label{eq:Sout}
\end{equation}
where $G$ is the gain of the measurement chain~\cite{Mallet:2011}. These fits yield three values of $N_{add}$ plotted as measurement efficiency
\begin{equation}
\eta_{m}=(2N_{add} + 1)^{-1} \label{eq:eta}
\end{equation}
\cite{Leonhardt:1994} in Fig.~\ref{fig:results}(c) (circles).

To find  $N_{add}$ at other values of the JPA gain, we use a model that decomposes $N_{add}$ into contributions from the JPA itself $N_{\mathrm{JPA}}$ and from the remaining measurement $N_{\mathrm{HEMT}}$. Adopting an added noise model, we interpolate $\eta_{m}$ over the range of JPA gains used in this experiment. In this model we assume that $N_{\mathrm{JPA}}$ and $N_{\mathrm{HEMT}}$ are both constant when referenced to their respective amplifier inputs. The output power spectral density is then
\begin{equation}
S_{out} = G (S_{in} +  N_{\mathrm{JPA}} + \frac{N_{\mathrm{HEMT}}}{G_{\mathrm{JPA}}}), \label{eq:SoutJPA}
\end{equation}
where $G_{\mathrm{JPA}}$ is the gain of the JPA. From Eq. \ref{eq:SoutJPA} the added noise can be written as
\begin{equation}
N_{add} = N_{\mathrm{JPA}} + \frac{N_{\mathrm{HEMT}}}{G_{\mathrm{JPA}}}, \label{eq:Nadd}
\end{equation}
forming the model we will use for interpolation~\cite{Beltran:2008}. The three added noises are fit according to Eq.~\ref{eq:Nadd}. We find $N_{\mathrm{JPA}} = 0.39\pm0.03$ and $N_{\mathrm{HEMT}} = 18 \pm 5$.   Using these added noises, $\eta_{m}$ is plotted over the range of JPA gains in Fig.~\ref{fig:results}(c) (solid line) according to Eqs.~\ref{eq:eta} and~\ref{eq:Nadd}.

\begin{acknowledgments}
We would like to thank Martin Sandberg, Michael Vissers, Jiansong Gao and David Pappas at NIST Boulder for providing the qubit used in the experiment, and James Thompson for an informative discussion. This work was supported by the National Science Foundation under grant number 1125844 and by the Gordon and Betty Moore Foundation.
\end{acknowledgments}

\bibliography{BibPhoGenKindel20150925v2}

\end{document}